\documentclass[article,12pt,3p]{elsarticle}
\usepackage{balance}
\usepackage{gensymb}
\usepackage{eurosym}
\usepackage{graphicx}
\usepackage{subfig}
\usepackage{amsmath}
\usepackage{mathtools}
\usepackage{amssymb}
\usepackage{siunitx}
\usepackage{booktabs}
\usepackage{arydshln}
\usepackage{algorithm}
\usepackage{algorithmic}
\usepackage{eurosym}
\usepackage{xr}
\usepackage{url}
\usepackage{xpunctuate}

\providecommand{\ie}[0]{\textit{i.e}\xperiod}
\providecommand{\etal}[0]{\textit{et al}\xperiod}

\usepackage[acronym, automake]{glossaries}
\newacronym{dsm}{DSM}{Demand Side Management}
\newacronym{dr}{DR}{Demand Response}
\newacronym{res}{RES}{Renewable Energy Sources}
\newacronym{hvac}{HVAC}{Heating, Ventilation and Air Conditioning}
\newacronym{bems}{BEMS}{Building Energy Management System}
\newacronym{ems}{EMS}{Energy Management System}
\newacronym{dlc}{DLC}{Direct Load Control}
\newacronym{ev}{EV}{Electric Vehicle}
\newacronym{pv}{PV}{photovoltaic}
\newacronym{fmu}{FMU}{Functional Mock-up Unit}
\newacronym{fmi}{FMI}{Functional Mock-up Interface}
\newacronym{ideas}{IDEAS}{Integrated District Energy Assessment Simulations}
\newacronym{gf}{gF}{Ground Floor}
\newacronym{cop}{COP}{Coefficient of Performance}
\newacronym{tcl}{TCL}{Thermostatically Controlled Load}
\newacronym{dso}{DSO}{Distribution System Operator}
\newacronym{dno}{DNO}{Distribution Network Operator}
\newacronym{ewh}{EWH}{Electric Water Heater}
\newacronym{dhw}{DHW}{Domestic Hot Water}
\newacronym{soc}{SoC}{State of Charge}
\newacronym{tou}{ToU}{Time-of-Use}
\newacronym{rdr}{RDR}{Residential Demand Response}
\newacronym{tso}{TSO}{Transmission System Operator}
\newacronym{mmp}{MMP}{Mean Month Peak}
\newacronym{sc}{SC}{Self-Consumption}
\newacronym{hems}{HEMS}{Home Energy Management System}

\newacronym{rl}{RL}{Reinforcement Learning}
\newacronym[firstplural=Markov Decision Processes (MDPs)]{mdp}{MDP}{Markov Decision Process}
\newacronym[firstplural=Partially Observable Markov Decision Processes (POMDPs)]{pomdp}{POMDP}{Partially Observable Markov Decision Process}
\newacronym{ml}{ML}{Machine Learning}
\newacronym{nn}{NN}{Neural Network}
\newacronym{dqn}{DQN}{Deep Q-network}
\newacronym{ppo}{PPO}{Proximal Policy Iteration}
\newacronym{gae}{GAE}{General Advantage Estimation}
\newacronym{drl}{DRL}{Deep Reinforcement Learning}
\newacronym{ai}{AI}{Artificial Intelligence}
\newacronym{ce}{CE}{Cross Entropy}
\newacronym{pdf}{PDF}{Probability Density Function}
\newacronym{mpc}{MPC}{Model Predictive Control}
\newacronym{er}{ER}{Experience Replay}
\newacronym{nfq}{NFQ}{Neural Fitted Q Iteration}
\newacronym{fqi}{FQI}{Fitted Q-Iteration}
\newacronym{dql}{DQL}{Double Q-learning}
\newacronym{mafqi}{MAFQI}{Model-Assisted Fitted Q-Iteration}
\newacronym{ann}{ANN}{Artificial Neural Network}
\newacronym{brl}{BRL}{Batch Reinforcement Learning}
\newacronym{cnn}{CNN}{Convolutional Neural Network}
\newacronym{svm}{SVM}{Support Vector Machine}
\newacronym{sdp}{SDP}{Sequential Decision making Problem}
\newacronym{milp}{MILP}{Mixed Integer Linear Problem}
\newacronym{rbc}{RBC}{Rule-Based Control}
\newacronym{hc}{HC}{Hysteresis Control}

\newacronym{us}{US}{United States}
\newacronym{wp}{WP}{Work Package}
\newacronym{eu}{EU}{European Union}
\newacronym{vreg}{VREG}{Vlaamse Regulator van de Energie- en Gasmarkt}

\glsmoveentry{us}{ign}
\glsmoveentry{eu}{ign}

\newglossaryentry{tasked}{
	name = \ensuremath{T_{\text{asked}}} ,
	description = Asked output water temperature,
	sort = tasked,
	symbol={\si{\degreeCelsius}}
}
\newglossaryentry{tsupply}{
	name = \ensuremath{T_{\text{supply}}} ,
	description = Supply water temperature,
	sort = tsupply
}
\newglossaryentry{tmax}{
	name = \ensuremath{T_{\text{max}}} ,
	description = {Maximum water temperature},
	sort = tmax,
	symbol={\si{\degreeCelsius}}
}
\newglossaryentry{tmin}{
	name = \ensuremath{T_{\text{min}}} ,
	description = Minimal buffer temperature,
	sort = tmin,
	symbol={\si{\degreeCelsius}}
}
\newglossaryentry{tbackup}{
	name = \ensuremath{T_b} ,
	description = Backup temperature measurement,
	sort = tb,
	symbol={\si{\degreeCelsius}}
}
\newglossaryentry{tenv}{
	name = \ensuremath{T_{\text{env}}} ,
	description = Environment temperature,
	sort = tenv
}
\newglossaryentry{tout}{
	name = \ensuremath{T_{\text{out}}} ,
	description = Outlet water temperature,
	sort = tout
}
\newglossaryentry{uphys}{
	name = \ensuremath{u_{\text{phys}}} ,
	description = Power after backup controller,
	sort = uphys
}
\newglossaryentry{discountrate}{
  name = \ensuremath{\gamma} ,
  description = The discount rate [-],
  sort = gamma
}
\newglossaryentry{state_set}{
  name = \ensuremath{\mathcal{X}} ,
  description = Set of states (state-space),
  sort = state
}
\newglossaryentry{state_set_obs}{
  name = \ensuremath{\mathcal{X}^o} ,
  description = Set of observable states (state-space),
  sort = stateobs
}
\newglossaryentry{state_set_markov}{
  name = \ensuremath{\mathcal{X}^a} ,
  description = Extended state space\, approximately possessing the Markov property,
  sort = statemarkov
}
\newglossaryentry{action_set}{
  name = \ensuremath{\mathcal{U}} ,
  description = Set of actions (action-space),
  sort = action
}
\newglossaryentry{conset}{
  name = \ensuremath{\mathcal{X}_c} ,
  description = Controllable component of the state-space,
  sort = xcon
}
\newglossaryentry{egset}{
  name = \ensuremath{\mathcal{X}_{eg}} ,
  description = Uncontrollable component of the state-space,
  sort = xeg
}
\newglossaryentry{timeset}{
  name = \ensuremath{\mathcal{X}_{T}} ,
  description = Time component of the state-space,
  sort = xtime
}
\newglossaryentry{egsetph}{
  name = \ensuremath{\mathcal{X}_{eg}^{ph}} ,
  description = Uncontrollable physical component of the state-space,
  sort = xegph
}
\newglossaryentry{egsetphforc}{
  name = \ensuremath{\hat{\mathcal{X}}_{eg}^{ph}} ,
  description = Forcast of the uncontrollable physical component of the state-space,
  sort = xegphfor
}
\newglossaryentry{egsetc}{
  name = \ensuremath{\mathcal{X}_{eg}^{c}} ,
  description = Uncontrollable external component of the state-space,
  sort = xegc
}
\newglossaryentry{policy}{
  name = \ensuremath{\pi} ,
  description = Policy: mapping from state-space to action-space,
  sort = policy
}
\newglossaryentry{rewardfunc}{
  name = \ensuremath{r(s_t, a_t, s_{t+1})} ,
  description = Reward function of Markov Decision Process,
  sort = reward
}
\newglossaryentry{valuefunc}{
  name = \ensuremath{V^\pi(s)} ,
  description = Value function,
  sort = valuefunc
}
\newglossaryentry{qfunc}{
  name = \ensuremath{Q(x,u)} ,
  description = Q-function in function of state x and action u,
  sort = qfunc
}
\newglossaryentry{optimalq}{
  name = \ensuremath{Q^*(s,a)} ,
  description = Optimal Q-function,
  sort = qfuncopt
}
\newglossaryentry{approxoptimalq}{
  name = \ensuremath{\hat{Q}^*(s,a)} ,
  description = Approximation of optimal Q-function,
  sort = qfuncoptapprox
}
\newglossaryentry{tsensor}{
  name = \ensuremath{T_{\text{sensor}}} ,
  description = The output temperature of the thermostat inside the building [\degree C],
  sort = tsensor
}
\newglossaryentry{temi}{
  name = \ensuremath{T_{\text{em}}} ,
  description = The temperature of the water leaving the emission system [\degree C],
  sort = temi
}
\newglossaryentry{irr}{
  name = \ensuremath{S_{\text{irr}}} ,
  description = The solar irradiance [W/$m^2$],
  sort = sirr
}
\newglossaryentry{qrad}{
  name = \ensuremath{Q_{rad}} ,
  description = Radiative heat power [W],
  sort = qrad
}
\newglossaryentry{qconv}{
  name = \ensuremath{Q_{conv}} ,
  description = Convective heat power [W],
  sort = qconv
}
\newglossaryentry{tset}{
  name = \ensuremath{T_{\text{set}}} ,
  description = Set-point temperature of the building [\degree C],
  sort = tset
}
\newglossaryentry{Tev}{
  name = \ensuremath{T_{\text{ev}}} ,
  description = Evaporator temperature of the heat pump [\degree C],
  sort = tev
}
\newglossaryentry{Tcon}{
  name = \ensuremath{T_{\text{con}}} ,
  description = Condenser temperature of the heat pump [\degree C],
  sort = tcon
}
\newglossaryentry{qasked}{
  name = \ensuremath{Q_{asked}} ,
  description = Asked thermal power from the heat pump [W],
  sort = qasked
}
\newglossaryentry{etamod}{
  name = \ensuremath{\eta_{mod}} ,
  description = Modulation degree of the heat pump [-],
  sort = etamod
}
\newglossaryentry{transprob}{
  name = \ensuremath{P_{x_tx_{t+1}}^{u_t}} ,
  description = Transition probability in a Markov Decision Process,
  sort = probab
}
\newglossaryentry{ftrans}{
  name = \ensuremath{f_T(x_t, u_{t}, P_{x_tx_{t+1}}^{u_t})} ,
  description = Transition function of Markov Decision Process,
  sort = ftrans
}
\newglossaryentry{epsgreedy}{
  name = \ensuremath{\epsilon} ,
  description = Probability of random action in epsilon-greedy policy,
  sort = epsgreedy
}
\newglossaryentry{costfunc}{
  name = \ensuremath{c(x_t, u_t, x_{t+1})} ,
  description = Cost function of the Markov Decision Process,
  sort = costfunc
}
\newglossaryentry{price}{
  name = \ensuremath{\lambda_t} ,
  description = Day-ahead electricity price at time t [\euro / MWh],
  sort = lambda
}
\newglossaryentry{batch}{
  name = \ensuremath{\mathcal{F}} ,
  description = Batch of transitions used in Batch RL,
  sort = f
}
\newglossaryentry{hppower}{
  name = \ensuremath{P_{el}} ,
  description = Electrical power consumed by the heat pump [W],
  sort = pel
}
\newglossaryentry{nnparam}{
	name = \ensuremath{\theta} ,
	description = Parameters of the neural network,
	sort = theta
}
\newglossaryentry{reals}{
	name = {real number},
	symbol = {\ensuremath{\mathbb{R}}},
	text = {\ensuremath{\mathbb{R}}},
	description = {Real numbers}
}
\newglossaryentry{boilercap}{
	name={\ensuremath{C_{\text{boiler}}}},
	description={Buffer capacitance},
	symbol={\si{\joule/\kelvin}}
}
\newglossaryentry{watercap}{
	name={\ensuremath{C_{pw}}},
	description={Water specific heat capacitance},
	symbol={\si{\joule/(\kilogram\kelvin)}}
}
\newglossaryentry{boilercond}{
	name={\textit{U}},
	description={Buffer thermal transmittance},
	symbol={\si{\watt/(\square\meter\kelvin)}}
}
\newglossaryentry{boilervol}{
	name={\ensuremath{V}},
	description={Buffer volume},
	symbol={\si{\kilogram}}
}
\newglossaryentry{boilersidesurf}{
	name={\ensuremath{A_s}},
	description={Buffer side surface},
	symbol={\si{\square\meter}}
}
\newglossaryentry{boilerheight}{
	name={\ensuremath{h}},
	description={Buffer height},
	symbol={\si{\meter}}
}
\newglossaryentry{boilerdiam}{
	name={\ensuremath{d}},
	description={Buffer diameter},
	symbol={\si{\meter}}
}
\newglossaryentry{boilertopsurf}{
	name={\ensuremath{A_t}},
	description={Buffer top/bottom surface},
	symbol={\si{\square\meter}}
}
\newglossaryentry{tapdemand}{
	name={\ensuremath{\dot{m}_w}},
	description={Hot water tap demand},
	symbol={\si{\kilogram/\second}}
}
\newglossaryentry{tinlet}{
	name={\ensuremath{T_{iw}}},
	description={Inlet water temperature},
	symbol={\si{\degreeCelsius}}
}
\newglossaryentry{tamb}{
	name={\ensuremath{T_{\text{amb}}}},
	description={Ambient temperature},
	symbol={\si{\degreeCelsius}}
}
\newglossaryentry{ratedpower}{
	name={\ensuremath{P_r}},
	description={Rated power},
	symbol={\si{\watt}}
}
\newglossaryentry{qheat}{
	name={\ensuremath{\dot{Q}_\text{heat}}},
	description={Heat flow rate heating element},
	symbol={\si{\joule/\second}}
}
\newglossaryentry{tlayer}{
	name={{\ensuremath{T_L}}},
	description={Water temperature (of layer i)},
	symbol={\si{\degreeCelsius}}
}
\newglossaryentry{optimalpolicy}{
	name={\ensuremath{\pi^*(x)}},
	description={Optimal policy}
}
\newglossaryentry{capprice}{
	name={\ensuremath{\lambda_P}},
	description={Capacity tariff},
	symbol={Euro/\si{\kilo\watt}}
}
\newglossaryentry{pnet}{
	name={\ensuremath{P_{\text{net}}}},
	description={Net power},
	symbol={\si{\watt}}
}
\newglossaryentry{domain}{
  name = \ensuremath{\mathcal{D}} ,
  description = Domain data,
  sort = domain
}

\makeglossaries

\usepackage{pgfplots, pgfplotstable}
\pgfplotsset{width=7cm,compat=1.8}
\definecolor{airforceblue}{rgb}{0.36, 0.54, 0.66}
\definecolor{antiquebrass}{rgb}{0.8, 0.58, 0.46}

\title{Combined Peak Reduction and Self-Consumption \\ Using Proximal Policy Optimization}

\author[1,2,3]{Thijs Peirelinck\corref{cor1}}%
\ead{thijs.peirelinck@kuleuven.be}

\author[2,3]{Chris Hermans}
\ead{chris.hermans@vito.be}

\author[2,3]{Fred Spiessens}
\ead{fred.spiessens@vito.be}

\author[1,3]{Geert Deconinck}
\ead{geert.deconinck@kuleuven.be}

\cortext[cor1]{Corresponding author}
\address[1]{ESAT, KU Leuven, Kasteelpark Arenberg 10 bus 2445, 3001 Leuven, Belgium}
\address[2]{AMO, Flemish Institute for Technological Research (VITO), Boeretang 200, 2400 Mol, Belgium}
\address[3]{AMO, EnergyVille, Thor Park 8310, 3600 Genk, Belgium}

\date{}

\begin{document}

\begin{abstract}
Residential demand response programs aim to activate demand flexibility at the household level. In recent years, reinforcement learning (RL) has gained significant attention for these type of applications. A major challenge of RL algorithms is data efficiency. New RL algorithms, such as proximal policy optimisation (PPO), have tried to increase data efficiency. Additionally, combining RL with transfer learning has been proposed in an effort to mitigate this challenge. In this work, we further improve upon state-of-the-art transfer learning performance by incorporating demand response domain knowledge into the learning pipeline. We evaluate our approach on a demand response use case where peak shaving and self-consumption is incentivised by means of a capacity tariff. We show our adapted version of PPO, combined with transfer learning, reduces cost by $14.51$\% compared to a regular hysteresis controller and by $6.68$\% compared to traditional PPO.
\end{abstract}

\maketitle

\glsresetall

\section{Introduction}
\gls{res} are reshaping the energy sector's landscape. Due to their decentralised and intermittent nature, market and tariff designs are challenged. In the Flemish region of Belgium the energy regulator (VREG) has recently announced a change of distribution fee design \cite{VREG2019}. Previously, Flemish residential electricity distribution fees have been energy-based. The rise of residential \gls{pv} installations and net-metering meant a reduction in income for the \gls{dno}. With the introduction of digital metering, the regulator takes the opportunity to introduce a capacity tariff, starting from 2023 \cite{VREG2019}. The regulator motivates its decision by arguing that the \gls{dno}'s main costs are capacity-based rather than energy-based \cite{VREG2019}. As a result, a more cost-reflective price signal is provided to consumers. At the same time, by tying the distribution grid fees to power rather than energy consumption, the VREG encourages consumers, aiming to minizime cost, to reduce their peak power consumption. This causes an interesting application for \gls{dr}, which we will consider in this work.

\glspl{tcl} are considered excellent appliances for \gls{dr}, as they have an inherent energy buffer \cite{Peirelinck2018, Patyn2018a}. In other \gls{dr} applications with \glspl{tcl}, \gls{rl} has shown promising results. For instance, Ruelens \etal \cite{ruelens_residential_2017} showed that \gls{fqi} can reduce energy consumption cost of a heat pump by $19$\,\%, compared to a default controller, in an energy arbitrage scenario. And Mbuwir \etal \cite{Mbuwir2020} apply the same algorithm for local optimisation in their transactive control framework. Their methodology manages to reduce grid congestion using flexibility available in the microgrid's heat pumps.

Additionally, multiple examples exist of successful \gls{ewh} control with \gls{rl} \cite{ruelens_residential_2017, Patyn2018, Peirelinck2018, kazmi_gigawatt-hour_2018, Peirelinck2020b, Bahrami2021, Vazquez-Canteli2019a}.
Reducing peak power consumption can (in part) be accomplished by local \gls{pv} self-consumption. Self-consumption is a \gls{dr} application in itself, and earlier work has applied \gls{rl} for maximising residential self-consumption. Soares \etal \cite{Soares2019, Soares2020} use a model-based \gls{rl} algorithm to control residential batteries and heat pumps. In their field-test, they achieve a $68$\,\% average self-consumption rate. This means, on average, $68$\,\% of heat pump energy is covered by local \gls{pv} generation. In a similar field-test, using the same model-based \gls{rl} approach, but only scheduling the \gls{ewh} heat cycle, De Somer \etal \cite{DeSomer2017} manage to increase \gls{pv} self-consumption by $20$\,\%.

The capacity tariff \gls{dr} application, which is considered here, differs from Soares \etal \cite{Soares2019, Soares2020} and De Somer \etal \cite{DeSomer2017} as the goal is not to maximise self-consumption. Rather, it is to minimise peak power consumption.
In the past, we have already touched upon a capacity tariff scenario \cite{Peirelinck2019a}.
However, the capacity tariff treated here is different, and is as proposed by the Flemish regulator in Belgium, \ie, VREG \cite{VREG2019}. The \gls{dr} use case of peak power reduction is of interest in other regions as well \cite{Azuatalam2020}.

State-of-the-art \gls{rl} has improved over time. Since its introduction, policy gradient methods \cite{Sutton2000} have gradually gained interest. Compared to value iteration methods, the policy gradient approach uses a function approximator that explicitly represents the policy \cite{Sutton2000}.
To further improve the policy gradient, which is used to update the approximator of the policy, an estimate of the expected future reward can be used \cite{Konda1999}.
This is achieved by the introduction of a critic \cite{Konda1999}. These actor-critic methods thus combine the advantages of value iteration and policy iteration \cite{Konda1999}.
\gls{ppo} \cite{Schulman2017} is the latest introduced family of actor-critic methods and is now widely used in \gls{rl} research. It has been introduced in an effort to mitigate two of the main drawbacks of \gls{rl}: (1) data (in)effiency and (2) the (dis)ability to perform well in a variety of domains \cite{Schulman2017}.
Earlier research \cite{ruelens_residential_2017, Patyn2018, Liu2019, Vazquez-Canteli2019a, Nweye2022} has shown the benefits of model-free \gls{rl} in residential \gls{dr} applications. In short, due to the hetergenuous nature of the appliances and users, the problem quickly becomes intractable. Additionally, the two mentioned properties of \gls{ppo} are of particular interest for \gls{dr}. The first property (data effiency) is important because the end user prefers to have a well performing agent as soon as possible. The second property (perform well in a variety of domains) is important because the agent needs to be able to quickly adapt its policy to the current season, as hot water demand and \gls{pv} output varies throughout the seasons.
Consequently, we have opted to use \gls{ppo} in this work. Furthermore, because earlier work \cite{Peirelinck2022, Zhang2022} has shown transfer learning can further improve data effiency of \gls{rl} algorithms, we combine \gls{ppo} with transfer learning.

Thus, in this work, \gls{ppo} has been adapted for automated \gls{dr} when a consumer with an \gls{ewh} is billed, at least partly, based on quarter hourly peak power consumption.
The \gls{rl} controller's aim is to minimise final energy cost by turning the \gls{ewh} on or off. Cost can be minimised by avoiding energy consumption when other (inflexible) loads are already using power or by self-consuming locally generated \gls{pv} power. To achieve this goal, we use transfer learning to pre-train the agent based on readily available consumption \cite{Edwards2015, baetens_openideas_2015} and production \cite{Pfenninger2016} data.
This paper's main contribution is the presentation of a data-efficient model-free residental \gls{dr} algorithm and learning pipeline. It does so by proposing a method to incorporate expert knowledge and transfer learning in the learning process, and by showing the benefits of a state-space design that explicitly takes into account transfer learning. Additionally, this paper contributes to design of \gls{rl} algorithms which can be applied in dynamically changing environments by showing that the proposed \gls{rl}-agent adapts to the seasonality of the control problem.

We test our approach on data from real households, obtained from a field-test in the Netherlands \cite{DeSomer2017, VanGoch2017}. Although the use case considers the Flemish tariff design, the method is more generally applicable. Moreover, reducing peak power demand is a widespread \gls{dr} setting.

This paper has been divided in four sections. Section \ref{sec::problem_algo} gives a more detailed formulation of the capacity tariff design, formulates the \gls{mdp} and lays out the \gls{rl} algorithm and its modifications.
The paper then goes on to Section \ref{sec::simulations}, presenting the experiments and discussing their results. Finally, Section \ref{sec::conclusions} concludes this work and gives future work directions.

\section{Problem Formulation \& Algorithm}\label{sec::problem_algo}
The first part of this section elaborates on the capacity tariff, as designed by the Flemish electricity and gas regulator (VREG). The second part defines the \gls{sdp}, and formulates it as an \gls{mdp}. The final part presents the algorithm used to solve the \gls{mdp}.

\subsection{Tariff Design}
A current Flemish residential electricity bill approximately consists of three parts; one part the energy cost [\euro/kWh], a second part the distribution costs [\euro/kWh] and a third part taxes and levies (only partly dependent on energy or power consumption). In the remainder of this work, we assume this simplified decoupling of the Flemish electricity bill. Traditionally, residential consumers only have a Ferraris meter installed, which is limited to net metering of energy consumption. The introduction of residential \gls{pv} installations meant the \gls{dno} saw its income reduce, as so-called \textit{prosumers} have relatively low net energy consumption and, as mentioned, distribution costs are energy-based. To compensate for this loss, a \textit{prosumer-tariff} was introduced. This tariff is charged based on the power inverter capacity of the \gls{pv} installation [\euro/$\text{kW}_{\text{inverter}}$] \cite{VREG2019}.

With the introduction of digital metering comes the ability to measure electricity consumption and production separately, and have finer grained measurement points. Together with the observation that distribution grid investment cost is mainly tied to grid capacity (and not energy transported), the regulator opted for a capacity-based distribution fee, from $2023$ onwards. This approximately results in the second part of the earlier mentioned electricity bill being capacity-based [\euro/kW] \cite{VREG2019}. As such, a capacity fee provides a more cost-reflective price signal to end-consumers.

The peak power to calculate the bill is based on the quarter-hourly measurements of the digital meter. The capacity fee of a residential consumer will be calculated based on the running \gls{mmp} of the past $12$ months. It only takes into account net off-take, \ie, there is no capacity fee based on grid injection. Thus, assuming $P^m_t$ is the quarter-hourly power consumption time-series of the current month $m$ in kW, \ie, the digital meter output, and \gls{capprice} is the price per kW in euro, the capacity fee $F$ of a residential consumer is calculated by Eq. \eqref{eq::cap_fee}.
\begin{align}
  \text{\gls{mmp}} &= \frac{\sum_{i=m-12}^m\max(2.5,\max(P^i_t))}{12} \label{eq::mmp}\\
  F &= \gls{capprice} \cdot \text{\gls{mmp}} \label{eq::cap_fee}
\end{align}
Eq. \eqref{eq::mmp} implies a minimal capacity fee based on a monthly peak power consumption of $2.5$\,kW, as in the tariff design \cite{VREG2019}. The main aim of this work is to minimise the final energy bill, \ie, including the parts related to energy consumption and taxes. The total energy bill is calculated by Eq. \eqref{eq::cap_bill}.
\begin{equation}\label{eq::cap_bill}
  C_{\text{capacity}} = \lambda_E \cdot E + \gls{capprice} \cdot \text{\gls{mmp}} + \lambda_{\text{tax}}^E \cdot E + \lambda_{\text{tax}}
\end{equation}
with $\lambda_E$ the energy price, $E$ the total energy consumption of the considered year, $\lambda_{\text{tax}}^E$ the taxes charged based on energy consumption and $\lambda_{\text{tax}}$ the fixed taxes payable per year. Eq. \eqref{eq::cap_bill} has been used to judge agent performance.

In \gls{rl}, the reward function can be used to direct the agent to optimal parts of the solution space, based on expert knowledge. For example, here we know self-consumption of locally generated \gls{pv} will be beneficial for both reducing energy consumption cost and reducing the \gls{mmp}. Furthermore, $\lambda_{\text{tax}}$ is independent of \gls{mmp} and $E$. As a consequence, neither equation \eqref{eq::cap_fee} nor equation \eqref{eq::cap_bill} is used as the \gls{mdp}'s reward-function. The following part of this section formulates the control problem as \gls{mdp} by presenting the state-space, action-space and reward-function.

\subsection{Markov Decision Problem}
The problem is formulated as a discrete-time \gls{mdp} with time steps of length $\Delta t = 15$ minutes. The \gls{mdp} consists of state-space \gls{state_set}, action-space \gls{action_set}, reward-function $r: (\gls{state_set}, \gls{action_set})\rightarrow \gls{reals}$ and state-transition probabilities $p(\cdot|x,u)$, as modeled by the \gls{ewh} model. The agent is unaware of these transition probabilities. The agent's goal is to learn a policy $\pi: \gls{state_set}\rightarrow\gls{action_set}$ which maximises cumulative reward.

With the setting as mentioned in the previous section, the main objective of this work is to reduce peak power consumption. Intuitively, reducing peak power consumption goes hand in hand with increasing self-consumption.

Our reward-function aims to formalise this intuition.
Following the \gls{mdp} framework, this objective is translated to reward-function \eqref{eq::rewardfunc_ratio}.
\begin{equation}\label{eq::rewardfunc_ratio}
r(x_t,u_t) =
\begin{cases}
\min(P^c - P^{net}_t, 0) + P^{sc}_t  & P^{\text{\gls{ewh}}}_t \neq 0 \\
0     & P^{\text{\gls{ewh}}}_t = 0
\end{cases}
\end{equation}
Where $P^c$ is $2.5$\,kW, 
$P^{\text{\gls{ewh}}}_t$ is the electrical power consumed by the \gls{ewh},
$P^{net}_t$ is the net power consumption and
$P^{sc}_t$ is the self-consumed \gls{ewh} power, at quarter $t$.
Given $P^D_t$ is the household's other inflexible electrical energy demand and $P^{PV}_t$ is the electrical power produced by the \gls{pv} installation,
the net power consumption $P^{net}_t$ and the \gls{ewh} self-consumption $P^{sc}_t$ are calculated by equations \eqref{eq::net_power} and \eqref{eq::ewh_selfcons}, respectively.
\begin{align}
  P^{net}_t &= P^{\text{\gls{ewh}}}_t + P^D_t - P^{PV}_t \label{eq::net_power}\\
  P^{sc}_t &= \min(\max(0, P^{PV}_t - P^D_t), P^{\text{\gls{ewh}}}_t) \label{eq::ewh_selfcons}
\end{align}

An additional challenge considered here is the aim of reducing the need for extensive local \gls{pv} and demand forecasts. On top of that, the learned control policy will be applied to different residential buildings and households. Therefore, to facilitate policy transfer, the state-space is designed to be independent of environment parameters. For instance, the learned policy should preferably be independent of inverter capacity, as different households will have different \gls{pv} installations.
Preference for policy independence on environment parameters can be illustrated by imagining a simple policy that turns the \gls{ewh} on whenever \gls{pv} power production is above $2$\,kW.
When this policy would be transferred to a different household, this absolute number does not apply anymore. Moreover, even within one single household this number would have to change between seasons.

At time-step $t \in \{0, \dots, 95\}$, state $x_t \in \gls{state_set}$ is defined by \eqref{eq::state}, with $\mu^T_t$ the mean temperature inside the buffer at time-step $t$,
$\Delta_{T_b^t - \gls{tmin}}$ the difference between the sensor measurement and the minimal allowed water temperature, $t_{\cos}$ and $t_{\sin}$ the projection of the time-step on a circle \cite{Peirelinck2020b}. Based on other work and to restore the Markov property we have opted to use a history of three time steps for the mean temperature \cite{ruelens_residential_2017, mnih_playing_2013}.
\begin{equation}\label{eq::state}
  x_t = \{\mu_t^T, \mu^T_{t-1}, \dots, \mu^T_{t-3}, \Delta_{T^b_t - \gls{tmin}},
  F^E_{PV}, \frac{P^{PV}_t}{F^P_{PV}}, t_{\cos}, t_{\sin}, t\}
\end{equation}
There are two state features dependent on a forecast of the local \gls{pv} production: $F^E_{PV}$ and $F^P_{PV}$. They are defined by equations \eqref{eq::energy_forecast} and \eqref{eq::power_forecast}, respectively.
$F^E_{PV}$ is the forecast of the current day's energy consumption, scaled with a rough estimate of the maximally possible energy production given the inverter power $P_{\text{inv}}$. And, $F^P_{PV}$ is a forecast of the peak power production of that same day. The agent has no (explicit) information on local power demand.
\begin{align}
  F^E_{PV} &= \sum_{i =\lfloor t/96 \rfloor}^{\lfloor t/96\rfloor + 96} \frac{E^{PV}_i}{96/2 \cdot \Delta t \cdot P_{\text{inv}}}     \label{eq::energy_forecast} \\
  F^P_{PV} &= \max_{i =\lfloor t/96 \rfloor}^{\lfloor t/96\rfloor + 96}\left(\frac{E^{PV}_i}{\Delta t}\right)  \label{eq::power_forecast}
\end{align}

This work considers a binary action-space. Every quarter $t$, the agent chooses an action $u_t \in \gls{action_set} = \{0, 1\}$, turning the \gls{ewh} on or off. When temperature constraints are violated, the controller overrules the agent's action according to Eq. \eqref{eq::backup_controller}, with $T^b_t$ the temperature at the backup sensor location (halfway up the buffer's height), $\gls{tmin} = 45\degree C$ and $\gls{tmax} = 55\degree C$.
\begin{equation}\label{eq::backup_controller}
\gls{uphys}= H(T^b_t, u_t) =
\begin{cases}
1					& T^b_t \leq \gls{tmin} \\
u_t				& T^b_t > \gls{tmin} \text{ and } T^b_t < \gls{tmax} \\
0					& T^b_t \geq \gls{tmax}
\end{cases}
\end{equation}

A two-layer \gls{ewh} model, similar to earlier work \cite{DeSomer2017, Soares2019, Soares2020, Peirelinck2020b} is used as a virtual test-bed. The model is based on the heat balance equations. A more detailed presentation is given in \cite{Peirelinck2020b}.

\subsection{Algorithm}
We use \gls{ppo} \cite{Schulman2017} to obtain a stochastic policy, maximising expected total reward, given reward-function \eqref{eq::rewardfunc_ratio}. It is an actor-critic algorithm. The family of actor-critic algorithms has been introduced in effort to combine the advantages of both policy-iteration and value-iteration algorithms. The actor \textit{acts}, \ie, it decides which action to take, and is trained based on a policy-iteration approach. The critic informs the actor about the state value and how it should update its policy in the right direction. The critic is trained with a value-iteration algorithm. Both actor and critic are parametrised using a (separate) \gls{nn}, with parameters $\theta_{\text{actor}}$ and $\theta_{\text{critic}}$, respectively.

At every iteration $k$, the actor's objective $L_k^{\text{actor}}$ \eqref{eq::actor_obj} and the critic's objective $L_k^{\text{critic}}$ \eqref{eq::critic_obj}
are (approximately) minimised.
\begin{align}
  L_k^{\text{actor}}\! &=\! \hat{\mathbb{E}}_k\!\left[\min\!\left(g_k(\theta_{\text{actor}})\hat{A}_k,\! \text{clip}(g_k(\theta_{\text{actor}}),\! 1\!-\!\epsilon,\! 1\!+\!\epsilon)\hat{A}_k)\!\right)\!\right]
  \label{eq::actor_obj} \\
  L_k^{\text{critic}}\! &=\! \hat{\mathbb{E}}_k\!\left[\left(V_{\theta_k^{\text{critic}}}(x_k) - V_k^{\text{targ}}\right)^2\right] \label{eq::critic_obj}
\end{align}
By the use of these update rules, an estimate of the value-function $V(x)$ and policy $\pi(x)$ is obtained by the critic and actor, respectively.
In Eq. \eqref{eq::actor_obj}, $r_k(\theta)$ denotes the probability ratio and is defined by \eqref{eq::prob_ratio_ppo} \cite{Schulman2017}. $\hat{A}_k$ denotes an advantage estimate and is defined by \eqref{eq::gae} \cite{Schulman2016}, with $\gamma$ and $\lambda$ hyper-parameters and $\delta_t^V$ given by \eqref{eq::delta_gae}.
The clipping in update rule \eqref{eq::actor_obj} avoids destructively large policy updates \cite{Schulman2017}. We use a clipping value of $\epsilon = 0.2$, and $\gamma = \lambda = 0.99$.
\begin{align}
  g_k(\theta) &= \frac{\pi_{\theta}(u_k | x_k)}{\pi_{\theta_{\text{old}}}(u_k | x_k)}
  \label{eq::prob_ratio_ppo} \\
  \hat{A}_k &= \sum_{l=0}^{\infty}(\gamma\lambda) (\delta_{t+l}^V) \label{eq::gae} \\
  \delta_t^V &= r_t + \gamma V(x_t+1) - V(x_t) \label{eq::delta_gae}
\end{align}

To improve convergence speed and policy transfer capabilities the actor-\gls{nn} is tailored to the task at hand, \ie, expert knowledge is incorporated into the actor \gls{nn}'s design. The domain knowledge is included in such a way that it does not restrict the applicability to one household or one type of \gls{tcl}. The actor's \gls{nn} has been designed based on three main observations:
\begin{itemize}
  \item Time-steps close to each other have a similar policy.
  \item If the current net consumption is close to the forecasted maximum \gls{pv} output of the day, its quite likely beneficial to heat water.
  \item The backup controller affects the policy and can be incorporated into the actor.
\end{itemize}

\begin{figure}
\centering
\includegraphics[width=0.4\linewidth]{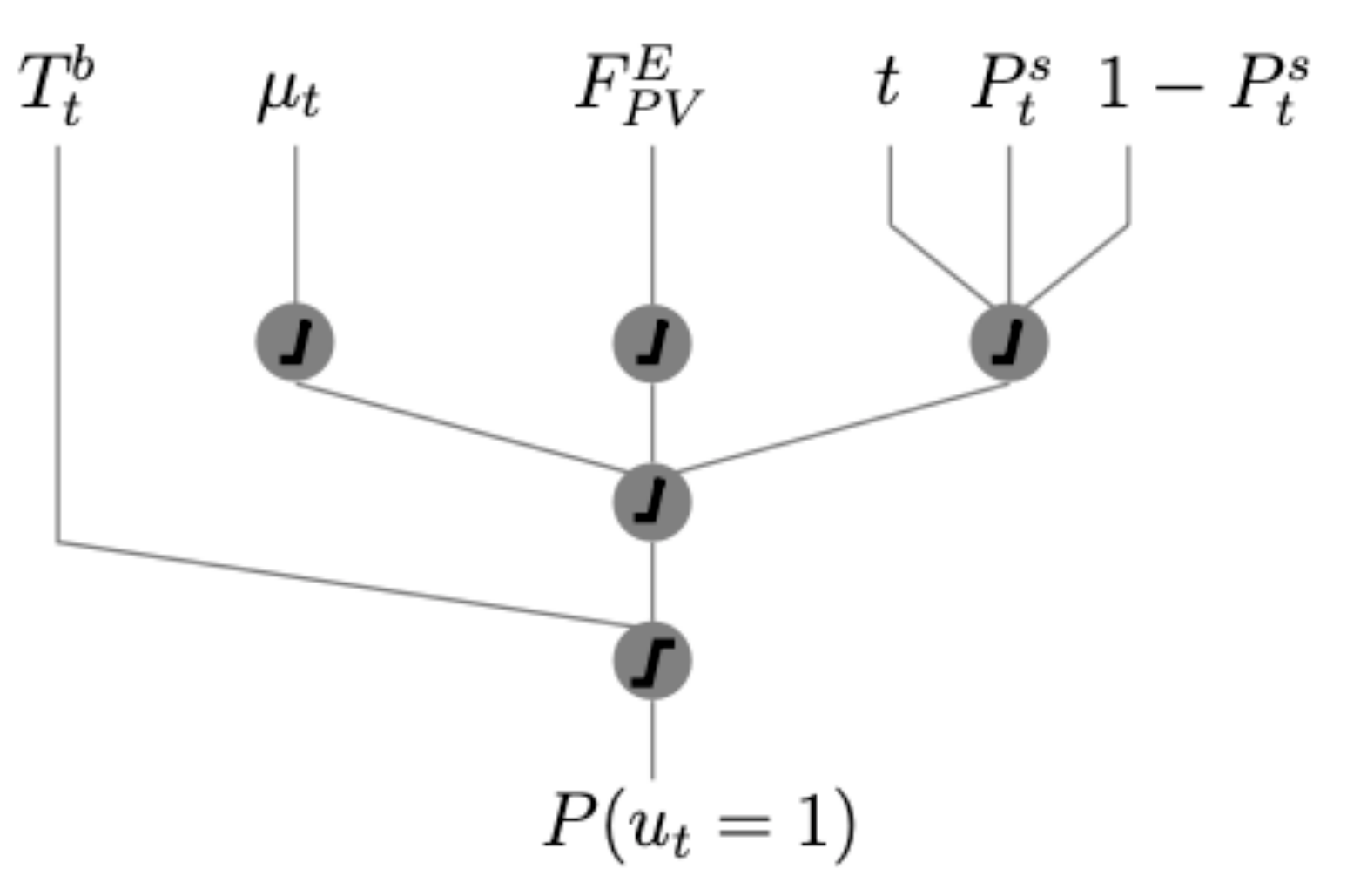}
\caption{Actor sub-NN architecture, with $P_t^s = P^{PV}_t/F^P_{PV}$.}
\label{fig::actor_nn}
\end{figure}

More specifically, the actor is divided in $24$ subnetworks, \ie, one for every hour of the day. Each of these networks has $3$ layers, of which the first is a feature extraction layer. The subnetwork architecture is shown in Fig.~\ref{fig::actor_nn}. The output of subnetwork $\mathcal{T} \in \{0, \dots, 23\}$, with parameters $\theta_{\mathcal{T}}^{\text{actor}}$, equals the probability of choosing action $u_t = 1$.
The final neuron of each subnetwork implements Eq. \eqref{eq::backup_controller}, assuring temperature stays within comfort bounds.
The actor thus uses only part of the observable state and, so does the critic. The full observable state is given in \eqref{eq::state}. The part used by the actor and critic is defined by \eqref{eq::state_actor} and \eqref{eq::state_critic}, respectively. The time-step is not omitted in $x_t^{\text{actor}}$, as it is needed to determine if it is better to turn the \gls{ewh} on at night or not.
\begin{align}
x_t^{\text{actor}} &= \{T_t^b, \mu_t^T, F^E_{PV}, \frac{P^{PV}_t}{F^P_{PV}}, t\} \label{eq::state_actor}\\
x_t^{\text{critic}} &= \{\mu_t^T, \mu^T_{t-1}, \dots, \mu^T_{t-3}, \Delta_{T^b_t - \gls{tmin}},
F^E_{PV}, \frac{P^{PV}_t}{F^P_{PV}}, t_{\cos}, t_{\sin}\} \label{eq::state_critic}
\end{align}
The critic's \gls{nn} has a conventional fully-connected architecture, with two layers of $28$ neurons. Apart from the neuron representing the backup controller, every neuron uses a ReLu activation function. All \glspl{nn} have been implemented in PyTorch \cite{paszke_automatic_2017}.

\section{Simulations \& Results}\label{sec::simulations}
\subsection{Experiment Set-up}
In earlier work we have shown transfer learning increases performance for agents applied in a \gls{dr} setting \cite{Peirelinck2020b}. Hence, the task is separated in a pre-training and test phase. The pre-training phase only uses readily available data. Three data streams are needed. First, simulated \gls{pv} production data is taken from the \textit{ninja} tool developed by Pfenninger and Staffell \cite{Pfenninger2016}. Second, electrical load data is generated using \textit{Strobe} \cite{baetens_openideas_2015}. Third, training phase simulations use \gls{dhw} consumption data from Edwards \etal \cite{Edwards2015}. These three sets contain data of one year and pre-training lasts $15$ simulation-years. After the initial pre-training phase, the obtained parameters $\theta_{\text{actor}}$ and $\theta_{\text{critic}}$
 are used as initial values for the test phase. During the test-phase, which lasts one simulation-year, we use real residential \gls{pv}, load and \gls{dhw} data, from five houses, obtained from a field-test in the Netherlands \cite{DeSomer2017, VanGoch2017}. Data is available starting from the first of October. Each experiment has been conducted $10$ times to account for variability in both phases. Table \ref{tab::metrics} gives some general metrics of the training- and test data. Clearly, a variety of households has been considered.

\begin{table}[]
\centering
\caption{General metrics of training- and test data, for 1 year.}
\label{tab::metrics}
\begin{tabular}{c|ccc}
         & DHW cons. [l/day] & $\sum E^D$ [kWh] & $\sum E^{PV}$ [kWh] \\ \hline
Training & $188.93$          & $3781.55$        & $3766.76$           \\ \hdashline
House 1  & $57.7$            & $2929.57$        & $7894.09$            \\
House 2  & $116.06$          & $5966.85$        & $8269.54$            \\
House 3  & $33.03$           & $3627.04$        & $7467.20$            \\
House 4  & $29.05$           & $3761.71$        & $8296.90$            \\
House 5  & $185.98$          & $5335.82$        & $7920.31$             \\
\end{tabular}
\end{table}

The presented approach has been compared with three other control approaches: \gls{hc}, \gls{rbc} and a non-expert version of \gls{rl} (\gls{ppo}). The hystersis (or, bang-bang) controller assures user comfort and turns the \gls{ewh} on or off according to Eq. \eqref{eq::backup_controller}. Like \gls{rl}, \gls{rbc} adds an additional layer on top of this hystersis controller. The implemented rule-based controller turns the \gls{ewh} on for four hours, at a fixed time $t_{\text{RBC}}$.
While the choice of $t_{\text{RBC}}$ may affect control performance, its optimal value is unknown beforehand.
Therefore, all \gls{rbc} simulations have been run four times, with $t_{\text{RBC}} \in \{10, 11, 12, 13\}$ hour. The non-expert version of \gls{rl} also uses \gls{ppo} as a training algorithm. However, the actor has a more traditional fully connected \gls{nn} with two layers of $10$ neurons each. This allows to confirm if the tailor made actor increases performance.

In the next section we show different result metrics, such as the final energy bill of the considered household, calculated by \eqref{eq::cap_bill}. The Flemish regulator has published capacity tariff values \gls{capprice} for different \glspl{dno}. We have chosen the average value $\gls{capprice} = 47.78$\,\euro/kW.

\subsection{Results}
This part presents the results of the simulations. We start with a visualisation of the three main control approaches (\gls{hc}, \gls{rbc}, (expert) \gls{rl}). Thereafter, we compare their performance differences in a more thorough manner. For the sake of simplicity, only expert \gls{rl} has been considered at the start.
It has been referred to as \gls{rl} from now on. Only at the final detailed comparison of the main metrics, non-expert \gls{rl} has been included.

\begin{figure}
  \centering
  \subfloat[House 3: 1 - 3 October ($t_{\text{RBC}} = 11$h).\label{fig::comp_days_2}]{\includegraphics[scale=1.]{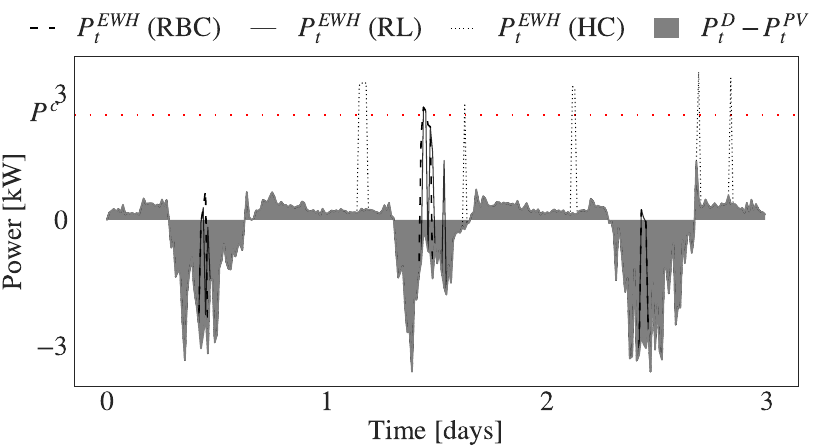}}\hfill
  \subfloat[House 1: 9 - 11 January ($t_{\text{RBC}} = 10$h).\label{fig::comp_days_0}]{\includegraphics[scale=1.]{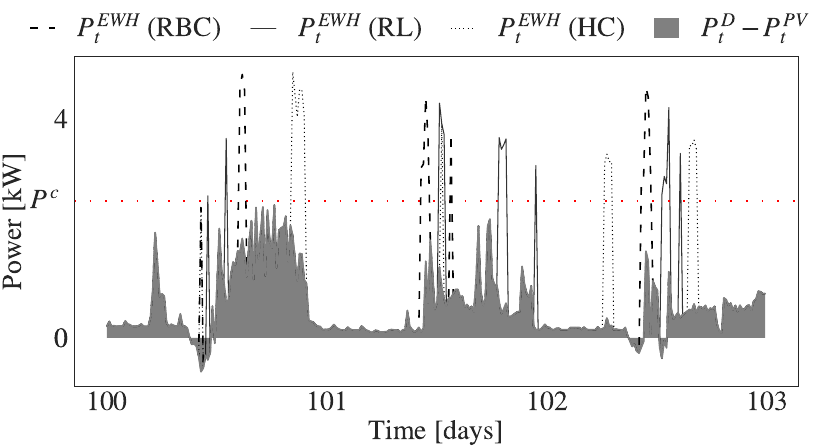}}\hfill
  \subfloat[House 2: 29 - 31 May ($t_{\text{RBC}} = 11$h).\label{fig::comp_days_1}]{\includegraphics[scale=1.]{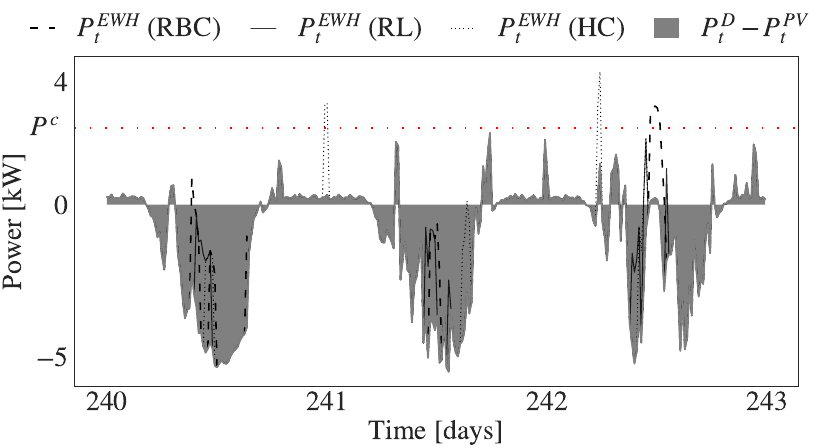}}\hfill
  \caption{Example days of three controllers and three houses, with best choice for $t_{\text{RBC}}$.}
  \label{fig::comp_days}
\end{figure}

Fig.~\ref{fig::comp_days} shows several test-phase days for the three considered control approaches. In each subfigure, the grey area depicts net uncontrollable load, \ie, inflexible load $P_t^D$ minus \gls{pv} production $P_t^{PV}$. The \gls{ewh}'s power demand $P_t^{\text{EWH}}$ for each control approach is depicted by different line-styles. As explained earlier, simulations have been conducted with several values for $t_{\text{RBC}}$.
In all subfigures of Fig.~\ref{fig::comp_days}, results of the $t_{\text{RBC}}$ value which resulted in the best final \gls{rbc} performance for the considered house has been shown.

Fig.~\ref{fig::comp_days_2} shows the three days immediately succeeding pre-training. At first sight, \gls{rbc} seems to be a good initial control approach, consuming power when local \gls{pv} production is high. This is, however, as expected, as the choice for \gls{rbc} is the result of expert domain knowledge. These three shown days further suggest \gls{rl} has resulted in similar behaviour as \gls{rbc}. This confirms the observation that \gls{rbc} is a rather good initial approach.

Fig.~\ref{fig::comp_days_0} shows three winter days, which illustrate how \gls{rl} further improves upon \gls{rbc}. The second and third day, the \gls{rl} agent turns the \gls{ewh} on at night, this avoids a heating cycle in the afternoon (during \gls{rbc} hours), when \gls{pv} production is low and consumption high. Clearly, \gls{hc} performs worse than the other approaches, as it does not take into account net consumption whatsoever.

Finally, Fig.~\ref{fig::comp_days_1} presents three spring days. All have quite a lot of \gls{pv} production and, therefore, clear negative consumption in the afternoon. However, \gls{pv} production is intermittent and has certain drops during the day. While the \gls{rl} agent has no forecast of \gls{pv} production in the next quarter, it has current net consumption as an input. As a result, and contrary to \gls{rbc}, it temporarily interrupts heating when \gls{pv} output suddenly drops. This is illustrated by the third day (day $242$) of Fig.~\ref{fig::comp_days_1}.

\begin{figure}
  \centering
  \includegraphics[width=0.7\linewidth]{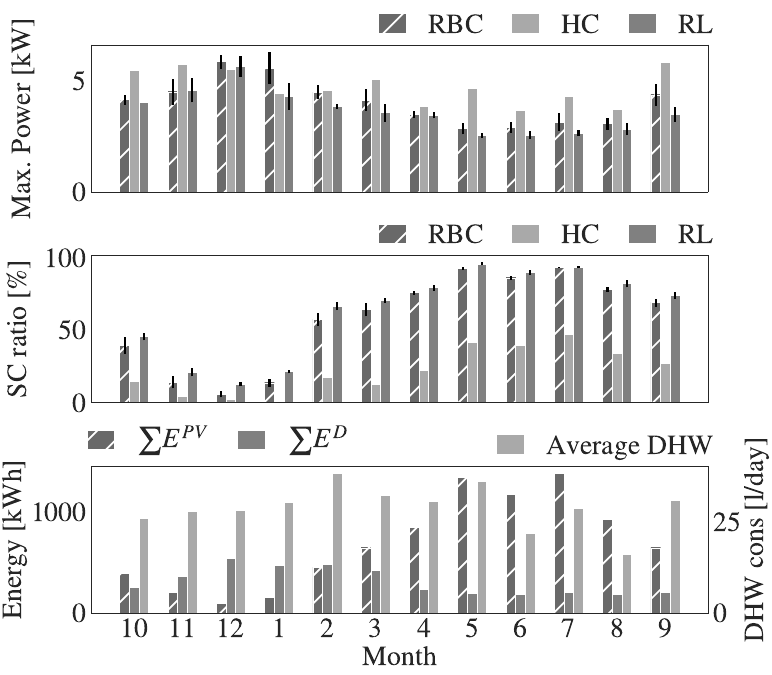}
  \caption{Test-year monthly performance metrics (house 4).}
  \label{fig::comp_year}
\end{figure}

Fig.~\ref{fig::comp_year} gives a more extensive overview of the results.
It presents an entire test-phase simulation year for house $3$.
The bottom graph shows each month's total \gls{pv} energy production $\sum E^{PV}$, total inflexible load $\sum E^D$ and average \gls{dhw} consumption.
These measures are useful for interpretation of control performance. Second, the middle graph shows self-consumption ratio of each month. For \gls{rbc} and \gls{rl} it shows mean and standard deviation of all simulation runs. The self-consumption ratio is the share of total \gls{ewh} power consumption which has been locally produced by the \gls{pv} installation. Intuitively, it is clear that this has to be maximised.
The graph shows \gls{rl} performs slightly better than \gls{rbc} in this regard, for all months. But, especially in months in which \gls{pv} production is low, \gls{rl} manages to capture more of the scarce local renewable energy for own consumption.
Finally, the top figure shows $P^{\text{max}}$, \ie, the month peak, for each month and for all three control approaches, with mean and standard deviation for \gls{rbc} and \gls{rl}. Remembering our final goal, $P^{\text{max}}$ should be minimised. \gls{rl} outperforms the other control approaches for all months, except November, in this household.

\begin{figure}
  \centering
  \includegraphics[width=0.7\linewidth]{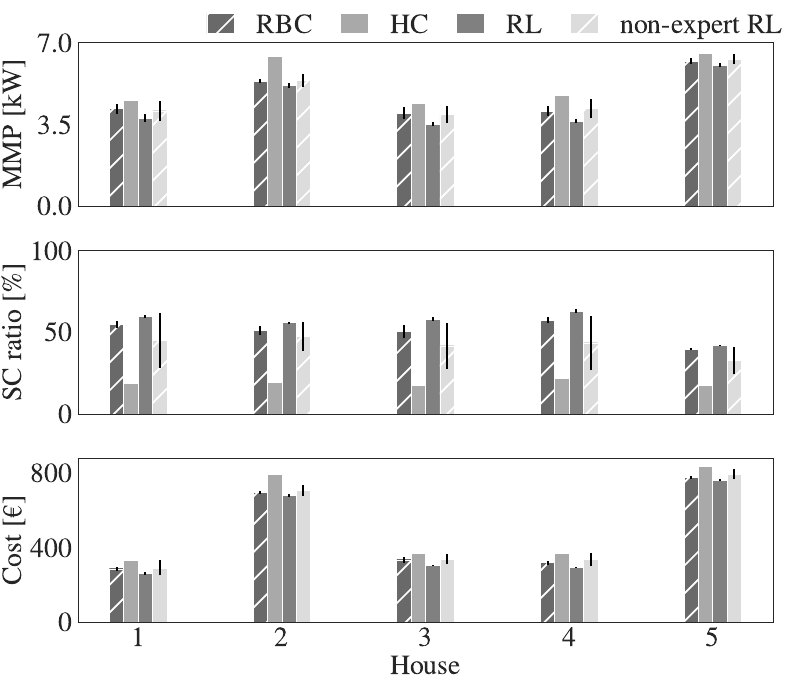}
  \caption{Final (test-phase) results for all houses.}
  \label{fig::comp_houses}
\end{figure}

In the end, the main goal is reducing the final yearly energy bills of households in Flanders, prone to a capacity tariff. Fig.~\ref{fig::comp_houses} shows all interesting metrics for the whole test-phase year and for all houses. The top figure shows the \gls{mmp}, calculated by \eqref{eq::mmp}. (Expert) \gls{rl} outperforms \gls{rbc}, \gls{hc} and non-expert \gls{rl} for each household.
More precisely, on average over all houses, \gls{rl} reduces the \gls{mmp} by $16.85$\,\% compared to \gls{hc}, and by $6.84$\,\% compared to \gls{rbc}.
In absolute numbers, this is a reduction of $0.90$\,kW and $0.33$\,kW, respectively.

As in Fig.~\ref{fig::comp_year}, the middle graph of Fig.~\ref{fig::comp_houses} shows the \gls{ewh}'s self-consumption ratio.
This figure shows that, for each household, \gls{rl} manages to shift \gls{ewh} power consumption better to quarters in which local energy production is available.
Compared to \gls{hc} and \gls{rbc}, expert \gls{rl} captures $192.34$\,\% and $9.93$\,\% more \gls{pv} production, respectively. This means that, over the year on average, $495.21$\,kWh more \gls{ewh} energy consumption is locally produced, compared to \gls{hc}, or $57.24$\,kWh compared to \gls{rbc}. Of further interest is that expert \gls{rl} manages to greatly reduce variability of the final performance, compared to non-expert \gls{rl}.

The bottom bar chart shows the final (electrical) energy bill of each household, defined by \eqref{eq::cap_bill}. (Expert) \gls{rl} is $14.51$\,\% cheaper than \gls{hc} and $4.59$\,\% cheaper than \gls{rbc}. For these households and the considered capacity cost, this thus results in an average reduction in cost of \euro$78.07$ and \euro$22.13$ compared to \gls{hc} and \gls{rbc}, respectively. Moreover, by incorporating domain knowledge into the \gls{rl} algorithm, we have managed to reduce costs with $6.68$\,\% or \euro$32.96$. Additionally, the performance's variability has decreased.

\begin{figure}
  \centering
  \includegraphics[width=0.7\linewidth]{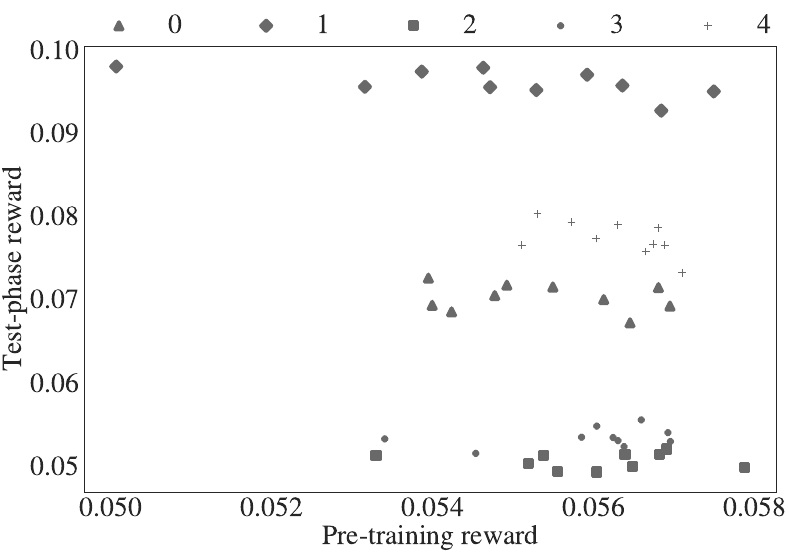}
  \caption{Scatter plot of mean pre-training-phase reward versus mean test-phase reward, for each of the five runs [0, 1, 2, 3, 4].}
  \label{fig::scatter_houses}
\end{figure}

The training-phase is an important step in the design and implementation of this \gls{rl} set-up for \gls{dr}. \gls{rl} is known to be data inefficient \cite{Schulman2017}. A pre-training-phase mitigates this drawbacks as data is less scarce in this phase. Fig.~\ref{fig::comp_days_2} has illustrated that, because of the pre-training-phase, the \gls{rl} agent performs its task well immediately at the start of the test-phase.
Fig.~\ref{fig::scatter_houses} aims to inspect if final pre-training performance affects test-phase performance. This figure shows that, although pre-training's final year mean reward varies between simulation runs, test-phase mean reward is always rather similar. This result suggests that, while it is important to pre-train the agent, it might not be necessary to somehow find the \textit{best} pre-trained agent. The average Pearson correlation coefficient between the mean reward of the final year of pre-training and the mean test-phase reward is $0.23$.

\section{Conclusions \& Future Work}\label{sec::conclusions}
We adapted state-of-art \gls{rl}, based on \gls{ppo}, to the \gls{dr} setting and have applied it for \gls{ewh} control in a capacity tariff use case. The considered setting is highly relevant in Belgium, as the (Flemish) regulator has decided to introduce a capacity tariff for residential consumers as of 2022.
In this scenario, the goal was identified to be two-fold; reduce the \gls{ewh}'s peak power consumption and increase self-consumption of local rooftop \gls{pv} production.
In our test-phase, we have used real-life data from five houses, all with different consumption patterns, but equipped with the same \gls{ewh}. Beforehand, the agent had been trained with readily available data. The setting is particularly challenging as the agent has no forecast of residential load and limited knowledge on future \gls{pv} production. Furthermore, we proposed an extension of \gls{ppo} where domain knowledge is incorporated into the actor's design. The results indicate that, using this approach, above human-level control performance is achieved.

In our experiments, we compared the \gls{rl} controller with \gls{rbc} and \gls{hc}. The experiments showed, for all considered houses, expert \gls{rl} outperforms both these two other control approaches. Self-consumption ratio was increased by $192.34$\,\% compared to \gls{hc}, and by $9.93$\,\% compared to \gls{rbc}. This resulted in a final energy bill reduction of $14.51$\,\% and $4.59$\,\% compared to \gls{hc} and \gls{rbc}, respectively. In absolute numbers, this translates to an average reduction of \euro$78.07$ and \euro$22.13$, respectively, in the yearly energy bill of the considered households.
One should note that, while \gls{rbc} performs relatively well, it is the result of expert knowledge and the best choice of $t_{\text{RBC}}$ differs for each house. Tuning this parameter needs expert knowledge, which comes at a cost. In contrary, once deployed, expert \gls{rl} automatically adapts to its environment. This has been shown in our experiments by the low correlation coefficient between pre-training and final mean reward.

Our experiments indicate that data which has been generated/collected for other puposes, such as district level simulations or feasability studies, can be used to improve \gls{rl} performance. This has the potential to facilitate \gls{rl}-based control uptake. Furthermore, while model-free learning has proven to be promising in \gls{dr} settings, it has also shown some drawbacks. Our experiments show that by including expert knowledge into the learning pipeline it is possible to mitigate some of these drawbacks, while still being generally applicable. This is in line with the current observations in physics-informed machine learning research.

We believe the introduction of a capacity tariff is an opportunity for Flemish residential consumers to adopt smart control approaches for \glspl{ewh} (and other \glspl{tcl}). Aggregators can potentially provide the necessary technology. Therefore, our future work is directed towards incorporating local control for reducing the \gls{mmp} within an aggregator framework.

\section{Acknowledgments}
\noindent Thijs Peirelinck, Chris Hermans and Fred Spiessens acknowledge support from the Flemish Institute for Technological Research (VITO) in the preparation of this manuscript.

%
\bibliographystyle{plain}
\bibliography{CapTariff}

\begin{thebibliography}{10}

\bibitem{Azuatalam2020}
Donald Azuatalam, Wee~Lih Lee, Frits de~Nijs, and Ariel Liebman.
\newblock {Reinforcement learning for whole-building HVAC control and demand
  response}.
\newblock {\em Energy and AI}, 2:100020, nov 2020.

\bibitem{baetens_openideas_2015}
Ruben Baetens, Roel {De Coninck}, Filip Jorissen, Damien Picard, Lieve Helsen,
  and Dirk Saelens.
\newblock {Openideas-an open framework for integrated district energy
  simulations}.
\newblock In {\em Proceedings of Building Simulation 2015}, pages 347 -- 354,
  2015.

\bibitem{Bahrami2021}
Shahab Bahrami, Yu~Christine Chen, and Vincent~W.S. Wong.
\newblock {Deep Reinforcement Learning for Demand Response in Distribution
  Networks}.
\newblock {\em IEEE Transactions on Smart Grid}, 12(2):1496--1506, mar 2021.

\bibitem{DeSomer2017}
Oscar {De Somer}, Ana Soares, Tristan Kuijpers, Koen Vossen, Koen Vanthournout,
  and Fred Spiessens.
\newblock {Using Reinforcement Learning for Demand Response of Domestic Hot
  Water Buffers: a Real-Life Demonstration}.
\newblock In {\em 2017 IEEE PES Innovative Smart Grid Technologies Europe (ISGT
  Europe)}, pages 1--7, 2017.

\bibitem{Edwards2015}
Skai Edwards, Ian Beausoleil-Morrison, and Andr{\'{e}} Laperri{\`{e}}re.
\newblock {Representative hot water draw profiles at high temporal resolution
  for simulating the performance of solar thermal systems}.
\newblock {\em Solar Energy}, 111:43--52, 2015.

\bibitem{kazmi_gigawatt-hour_2018}
Hussain Kazmi, Fahad Mehmood, Stefan Lodeweyckx, and Johan Driesen.
\newblock {Gigawatt-hour scale savings on a budget of zero: {Deep}
  reinforcement learning based optimal control of hot water systems}.
\newblock {\em Energy}, 144:159--168, 2018.

\bibitem{Konda1999}
Vijay~R Konda and John~N Tsitsiklis.
\newblock {Actor-Critic Algorithms}.
\newblock In {\em Advances in Neural Information Processing Systems 12 (NIPS
  1999)}, pages 1008--1014, 1999.

\bibitem{Liu2019}
Mingxi Liu, Stef Peeters, Duncan~S. Callaway, and Bert~J. Claessens.
\newblock {Trajectory Tracking With an Aggregation of Domestic Hot Water
  Heaters: Combining Model-Based and Model-Free Control in a Commercial
  Deployment}.
\newblock {\em IEEE Transactions on Smart Grid}, 10(5):5686--5695, 2019.

\bibitem{Mbuwir2020}
Brida~V. Mbuwir, Fred Spiessens, and Geert Deconinck.
\newblock {Distributed optimization for scheduling energy flows in community
  microgrids}.
\newblock {\em Electric Power Systems Research}, 187:106479, oct 2020.

\bibitem{mnih_playing_2013}
Volodymyr Mnih, Koray Kavukcuoglu, David Silver, Alex Graves, Ioannis
  Antonoglou, Daan Wierstra, and Martin Riedmiller.
\newblock {Playing Atari with Deep Reinforcement Learning}.
\newblock In {\em NIPS Deep Learning Workshop 2013}, 2013.

\bibitem{Nweye2022}
Kingsley Nweye, Bo~Liu, Peter Stone, and Zoltan Nagy.
\newblock {Real-world challenges for multi-agent reinforcement learning in
  grid-interactive buildings}.
\newblock {\em Energy and AI}, 10:100202, nov 2022.

\bibitem{paszke_automatic_2017}
Adam Paszke, Sam Gross, Francisco Massa, Adam Lerer, James {Bradbury Google},
  Gregory Chanan, Trevor Killeen, Zeming Lin, Natalia Gimelshein, Luca Antiga,
  Alban Desmaison, Andreas~K{\"{o}}pf Xamla, Edward Yang, Zach Devito, Martin
  {Raison Nabla}, Alykhan Tejani, Sasank Chilamkurthy, Qure Ai, Benoit Steiner,
  Lu~Fang Facebook, Junjie~Bai Facebook, and Soumith Chintala.
\newblock {PyTorch: An Imperative Style, High-Performance Deep Learning
  Library}.
\newblock In {\em Advances in Neural Information Processing Systems 32 (NIPS
  2019)}, pages 8026--8037. Curran Associates, Inc., 2019.

\bibitem{Patyn2018}
Christophe Patyn, Thijs Peirelinck, and Geert Deconinck.
\newblock {Intelligent Electric Water Heater Control with Varying State
  Information}.
\newblock In {\em 2018 IEEE International Conference on Communications,
  Control, and Computing Technologies for Smart Grids (SmartGridComm)}, pages
  1--7. IEEE, 2018.

\bibitem{Patyn2018a}
Christophe Patyn, Frederik Ruelens, and Geert Deconinck.
\newblock {Comparing neural architectures for demand response through
  model-free reinforcement learning for heat pump control}.
\newblock In {\em 2018 IEEE International Energy Conference, ENERGYCON 2018},
  pages 1--6. IEEE, 2018.

\bibitem{Peirelinck2020b}
Thijs Peirelinck, Chris Hermans, Fred Spiessens, and Geert Deconinck.
\newblock {Domain Randomization for Demand Response of an Electric Water
  Heater}.
\newblock {\em IEEE Transactions on Smart Grid}, 12(2):1370--1379, may 2020.

\bibitem{Peirelinck2022}
Thijs Peirelinck, Hussain Kazmi, Brida~V Mbuwir, Chris Hermans, Fred Spiessens,
  Johan Suykens, and Geert Deconinck.
\newblock {Transfer learning in demand response: A review of algorithms for
  data-efficient modelling and control}.
\newblock {\em Energy and AI}, 7:100126, 2022.

\bibitem{Peirelinck2018}
Thijs Peirelinck, Frederik Ruelens, and Geert Deconinck.
\newblock {Using reinforcement learning for optimizing heat pump control in a
  building model in Modelica}.
\newblock In {\em 2018 IEEE International Energy Conference (ENERGYCON)}, pages
  1--6. IEEE, 2018.

\bibitem{Peirelinck2019a}
Thijs Peirelinck, Fred Spiessens, Chris Hermans, and Geert Deconinck.
\newblock {Double Q-learning for Demand Response of an Electric Water Heater}.
\newblock In {\em 2019 IEEE PES Innovative Smart Grid Technologies Europe (ISGT
  Europe)}, pages 1--5. IEEE, 2019.

\bibitem{Pfenninger2016}
Stefan Pfenninger and Iain Staffell.
\newblock {Long-term patterns of European PV output using 30 years of validated
  hourly reanalysis and satellite data}.
\newblock {\em Energy}, 114:1251--1265, nov 2016.

\bibitem{ruelens_residential_2017}
F~Ruelens, B~J Claessens, S~Vandael, B~{De Schutter}, R~Babu{\v{s}}ka, and
  R~Belmans.
\newblock {Residential Demand Response of Thermostatically Controlled Loads
  Using Batch Reinforcement Learning}.
\newblock {\em IEEE Transactions on Smart Grid}, 8(5):2149--2159, sep 2017.

\bibitem{Schulman2016}
John Schulman, Philipp Moritz, Sergey Levine, Michael~I. Jordan, and Pieter
  Abbeel.
\newblock {High-dimensional continuous control using generalized advantage
  estimation}.
\newblock In {\em 4th International Conference on Learning Representations,
  ICLR 2016}, jun 2016.

\bibitem{Schulman2017}
John Schulman, Filip Wolski, Prafulla Dhariwal, Alec Radford, and Oleg Klimov.
\newblock {Proximal Policy Optimization Algorithms}.
\newblock jul 2017.

\bibitem{Soares2019}
Ana Soares, Oscar {De Somer}, Dominic Ectors, Filip Aben, Jan Goyvaerts, Milo
  Broekmans, Fred Spiessens, Dennis {Van Goch}, and Koen Vanthournout.
\newblock {Distributed Optimization Algorithm for Residential Flexibility
  Activation-Results from a Field Test}.
\newblock {\em IEEE Transactions on Power Systems}, 34(5):4119--4127, 2019.

\bibitem{Soares2020}
Ana Soares, Davy Geysen, Fred Spiessens, Dominic Ectors, Oscar {De Somer}, and
  Koen Vanthournout.
\newblock {Using reinforcement learning for maximizing residential
  self-consumption – Results from a field test}.
\newblock {\em Energy and Buildings}, 207, 2020.

\bibitem{Sutton2000}
Richard~S Sutton, David Mcallester, Satinder Singh, and Yishay Mansour.
\newblock {Policy Gradient Methods for Reinforcement Learning with Function
  Approximation}.
\newblock In {\em Advances in Neural Information Processing Systems (NIPS
  1999)}, volume~12, pages 1057--1063. MIT Press, 2000.

\bibitem{VanGoch2017}
Dennis van Goch, Marc Eulen, Stefan Lodeweyckx, and Chris Caerts.
\newblock {Rennovates, Flexibility Activated Zero Energy Districts, H2020}.
\newblock Technical report, 2017.

\bibitem{Vazquez-Canteli2019a}
Jos{\'{e}}~R. V{\'{a}}zquez-Canteli and Zolt{\'{a}}n Nagy.
\newblock {Reinforcement learning for demand response: A review of algorithms
  and modeling techniques}.
\newblock {\em Applied Energy}, 235:1072--1089, feb 2019.

\bibitem{VREG2019}
{Vlaamse Regulator Energie en Gas (VREG)}.
\newblock {Toekomst nettarieven – capaciteitstarief | VREG}.

\bibitem{Zhang2022}
Xiangyu Zhang, Yue Chen, Andrey Bernstein, Rohit Chintala, Peter Graf, Xin Jin,
  and David Biagioni.
\newblock {Two-Stage Reinforcement Learning Policy Search for Grid-Interactive
  Building Control}.
\newblock {\em IEEE Transactions on Smart Grid}, 13(3):1976--1987, may 2022.

\end{thebibliography}

\end{document}